%% file: FDD.tex
\makeatletter
\def\CLASSINPUTbottomtextmargin{1in}
\makeatother

\documentclass[conference]{IEEEtran}
\usepackage{cite}

\ifCLASSINFOpdf
  \usepackage[pdftex]{graphicx}
  \graphicspath{{../pdf/}{../jpeg/}}
  \DeclareGraphicsExtensions{.pdf,.jpeg,.png}
\else
\fi
\usepackage[cmex10]{amsmath}
\usepackage{amssymb}

\usepackage[caption=false,font=footnotesize]{subfig}

\usepackage{tikz,pgfplots}
\pgfplotsset{compat=1.18}
\usetikzlibrary{backgrounds}
\usetikzlibrary{arrows.meta}
\usetikzlibrary{positioning,fit,calc}
\tikzset{fit margins/.style={/tikz/afit/.cd,#1,
/tikz/.cd,
inner xsep=\pgfkeysvalueof{/tikz/afit/left}+\pgfkeysvalueof{/tikz/afit/right},
inner ysep=\pgfkeysvalueof{/tikz/afit/top}+\pgfkeysvalueof{/tikz/afit/bottom},
xshift=-\pgfkeysvalueof{/tikz/afit/left}+\pgfkeysvalueof{/tikz/afit/right},
yshift=-\pgfkeysvalueof{/tikz/afit/bottom}+\pgfkeysvalueof{/tikz/afit/top}},
afit/.cd,left/.initial=2pt,right/.initial=2pt,bottom/.initial=2pt,top/.initial=5pt}
\tikzset{base/.style={rectangle, rounded corners, draw=black, thick, text centered, fill=blue!20},
innerbase/.style={base, draw=black!30},
eebox/.style={base, dashed, thick, fill=black!08, inner xsep=0.5cm},
explotation/.style={base, fill=red!40!white},
eephase/.style={font=\large\bfseries},
connector/.style={very thick,-{Latex[width=2.5mm]}, rounded corners}
}

\begin{document}
\IEEEoverridecommandlockouts
%
\title{A Robust Method for Fault Detection and Severity Estimation in Mechanical Vibration Data}

\author{
\IEEEauthorblockN{Youngjae Jeon}
\IEEEauthorblockA{Department of Automotive\\ Engineering (Automotive-\\Computer Convergence)\\
Hanyang University\\
Seoul, South Korea\\
yjeon@hanyang.ac.kr}
\\ 
\IEEEauthorblockN{Dongjin Lee*}
\IEEEauthorblockA{Department of Automotive\\ Engineering\\
Hanyang University\\
Seoul, South Korea\\
dlee46@hanyang.ac.kr}
\and
\IEEEauthorblockN{Eunho Heo}
\IEEEauthorblockA{Department of Automotive\\ Engineering (Automotive-\\Computer Convergence)\\
Hanyang University\\
Seoul, South Korea\\
gjrkd426@hanyang.ac.kr}
\and
\IEEEauthorblockN{Jinmo Lee}
\IEEEauthorblockA{Airbility\\
Seoul, South Korea\\
jm.lee@airbility.co.kr}
\and
\IEEEauthorblockN{Taewon Uhm}
\IEEEauthorblockA{Airbility\\
Seoul, South Korea\\
taewon.uhm@airbility.co.kr}
}

%
%
%
%
%
%
%
\IEEEpubid{\begin{minipage}{\textwidth}
    \footnotesize * Corresponding author
\end{minipage}} 
\IEEEpubidadjcol 
%
\maketitle
\begin{abstract}
This paper proposes a robust method for fault detection and severity estimation in multivariate time-series data to enhance predictive maintenance of mechanical systems.
We use the Temporal Graph Convolutional Network (T-GCN) model to capture both spatial and temporal dependencies among variables. This enables accurate future state predictions under varying operational conditions.
To address the challenge of fluctuating anomaly scores that reduce fault severity estimation accuracy, we introduce a novel fault severity index based on the mean and standard deviation of anomaly scores. This generates a continuous and reliable severity measurement.
We validate the proposed method using two experimental datasets: an open IMS bearing dataset and data collected from a fanjet electric propulsion system. 
Results demonstrate that our method significantly reduces abrupt fluctuations and inconsistencies in anomaly scores. This provides a more dependable foundation for maintenance planning and risk management in safety-critical applications.
\end{abstract}


%
\IEEEpeerreviewmaketitle

\section{Introduction}  


Predictive maintenance serves as a critical tool in Prognostics and Health Management (PHM) for modern industries--particularly power generation and automotive systems--where there is an increasing demand for proactive maintenance strategies to prevent unexpected failures and optimize operational efficiency~\cite{chandola2009anomaly}. Fault detection and fault severity estimation, as primary components of predictive maintenance, enable identification of faults within the system. Fault detection determines both the occurrence and condition of faults~\cite{arunthavanathan2021analysis}, while fault severity estimation focuses on determining a severity index that quantifies the extent of damage~\cite{cerrada2018review}.

Fault detection methods fall into three main categories: model-based, knowledge-based, and data-driven~\cite{arunthavanathan2021analysis}. Machine learning methods, such as random forest, autoregressive integrated moving average model, and support vector machine--all part of the data-driven approach--identify fault patterns by extracting and learning features directly from data. However, these methods struggle to capture the complex patterns inherent in the data~\cite{xu2021two}.
%
The recent works~\cite{pang2021deep,lin2024vibration} present that deep learning and its variants effectively capture temporal dependencies and nonlinear relationships between variables in the data.  
Recurrent neural networks (RNNs), including variants like long short-term memory~(LSTM) and gated recurrent unit~(GRU), are used for predicting future states~\cite{malhotra2016lstm, nizam2022real, zhang2023structural, choi2022multivariate}.
Graph based networks such as graph neural network (GNN) and graph convolutional network (GCN) are used to capture complex spatial dependencies and extract features from graph-structured data~\cite{shi2023robust, xiao2023graph, liao2020fault, wang2020temporal}. 
Temporal graph neural network (T-GCN)~\cite{zhao2019t} combines GCN with GRU to effectively capture both temporal dependencies and nonlinear relationships among variables. In this work, we use T-GCN as it achieves robust fault detection in both nonlinear data and varying environmental conditions.

For fault severity estimation, the previous works~\cite{yu2023new, saeed2020novel, saeed2020online, li2019deep, husari2022stator, medina2019deep, xu2021two, luo2020rolling, li2016quantitative} use predefined defect levels that are heuristically determined to estimate fault severity. This approach makes defect levels highly dependent on individual human experience. The work~\cite{wang2024research} struggles to capture complex relationships between multiple variables, resulting in unstable assessments when anomaly scores fluctuate abruptly. To overcome these limitations, we need to develop a more robust fault severity estimation method. 



  



We present a robust fault diagnosis method that combines T-GCN with a novel severity estimation method in mechanical vibration data. T-GCN simultaneously considers the temporal and spatial dependencies among multiple output variables, and thus achieves robust future predictions across different operating conditions, such as location and temperature. We estimate the fault severity level of output variables by computing its mean and standard deviation over time. This method allows for more accurate fault prediction by reducing the impact of highly fluctuating state data. Finally, we demonstrate the robustness and prediction accuracy of the proposed method using experimental vibration datasets for bearing and fanjet electric propulsion systems. 
%

%
The paper is organized as follows. Section~\ref{background} covers the background on the deep learning model, fault detection, and fault severity estimation. Section~\ref{method} proposes a novel method for robust fault analysis and severity estimation method. Section~\ref{exp} demonstrates the robustness of the proposed method using bearing and fanjet datasets. Section~\ref{con} presents the conclusions and limitations of the work.


\section{Background} \label{background}

\subsection{Temporal Graph Convolution Network} 

\begin{figure}[!t] \vspace*{0.05in} \centering \includegraphics[width=2.4in]{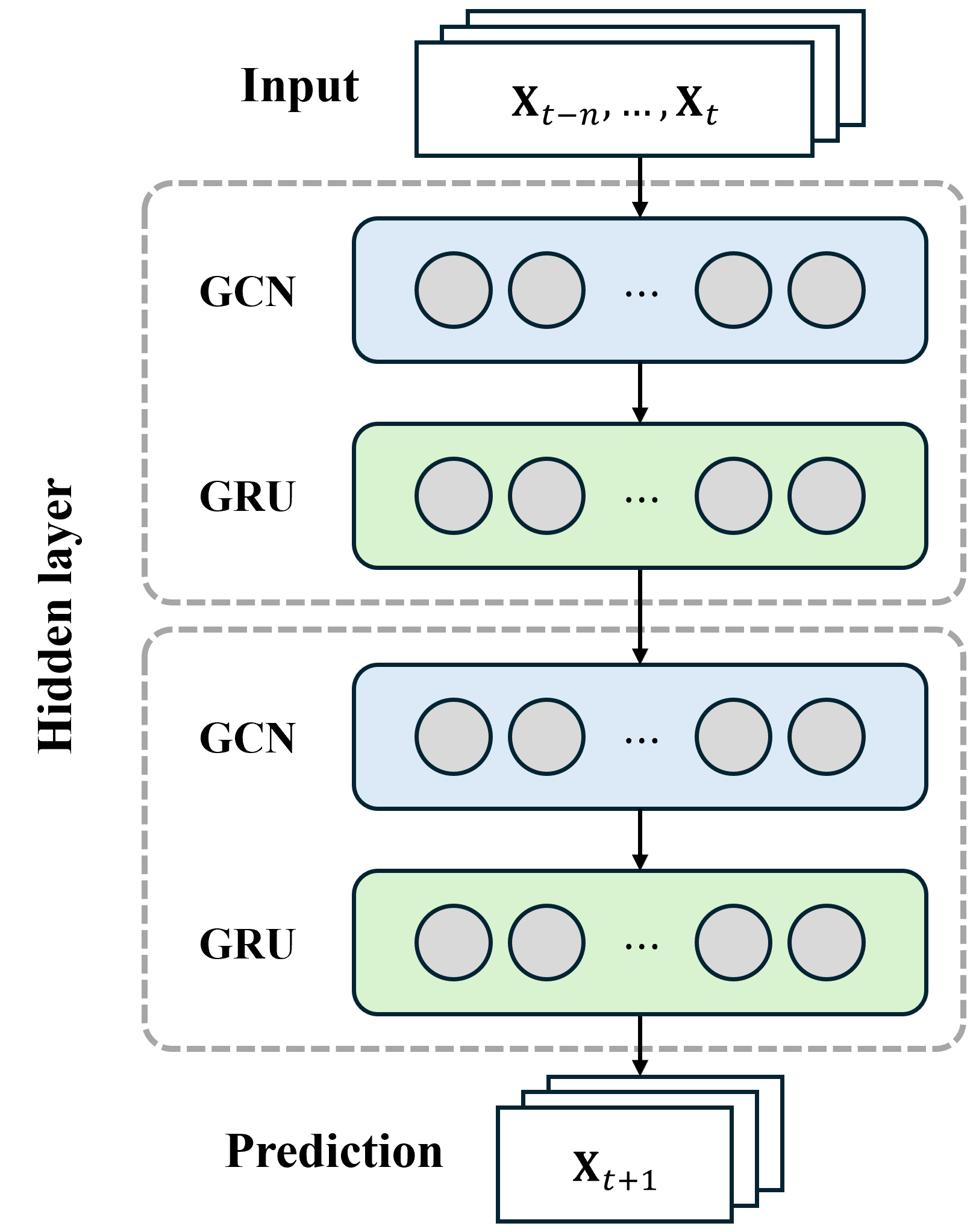} \caption{Schematic diagram of 2-layer T-GCN structure} \label{T-GCN} \end{figure}

T-GCN is a deep learning model for capturing spatial and temporal features. T-GCN consists of GCN and GRU as shown in Fig.\ref{T-GCN}. GCN captures spatial features in a graph structure reflecting relationship between variables. GRU captures temporal dependence using recurrent unit that can transfer temporal features.

Let us define graph as $\mathcal{G} = \{\mathcal{V}, \mathbf{A}\}$, where $\mathcal{V}$ is the set of nodes $v_i \in \mathcal{V},\quad i=1,\dots,N$, and $N$ is the number of the nodes.
%
The adjacency matrix $\mathbf{A} \in \mathbb{R}^{N\times N}$ represents the relationships among nodes. The entries of $\mathbf{A}$ are binary: $0$ indicates the absence of a connection between the corresponding nodes, while $1$ indicates the presence of a connection.
%
The feature matrix $\mathbf{X} = [\mathbf{X}_1, \dots, \mathbf{X}_T] \in \mathbb{R}^{N\times T}$ contains the temporal and spatial features of the nodes such as rotating speed and vibration magnitude. $T$ is the length of the time series data.

The spatial feature can be considered as training a mapping function on the adjacency matrix $\mathbf{A}$ and the feature matrix $\mathbf{X}$. The mapping function captures the spatial relations of the input data. 
%
Given the adjacency matrix $\mathbf{A}$ and the feature matrix, the convolutional layer can be represented as
\begin{align}
    \mathbf{H}^{(l+1)} = \sigma \left( \tilde {\mathbf{A}} \mathbf{H}^{(l)} \mathbf{W}^{(l)} \right),
\end{align}
%
where $\mathbf{H}^{(l+1)}$ and $\mathbf{H}^{(l)}$ are the output and the input node representations of $l$th layer, respectively. The initial node representation $\mathbf{H}^{(0)}$ is the input feature matrix $\mathbf{X}$. Sigmoid function $\sigma(\cdot)$ is for a nonlinear representation. $\mathbf{W}^{(l)}$ is the weight matrix of the layer. $\tilde{\mathbf{A}} = \mathbf{A} + \mathbf{I}$ is the self-connection added adjacency matrix, $\mathbf{I}$ is the identity matrix. 
%
Each node in the graph has a different number of neighbors. Nodes with fewer connections may have their features underrepresented compared to nodes with more connections, leading to potential distortions. To address this, the adjacency matrix is normalized to ensure a fair contribution of the characteristics based on the number of neighbors. The normalized adjacency matrix $\hat{\mathbf{A}}$ is obtained as
\begin{align}
    \hat {\mathbf A} = \tilde {\mathbf D}^{-1/2} \tilde {\mathbf A} \tilde {\mathbf D}^{-1/2},
\end{align}
%
where $\tilde {\mathbf D}$ is the degree matrix. The diagonal element $\tilde{\mathbf{D}}_{ii} = \sum_{j=1}^{N}{\tilde{\mathbf{A}}_{ij}}$ represents the number of neighbors for node $i$. Using the normalized adjacency matrix, the convolutional layer and the mapping function $f$ can be represented as 
%
\begin{align}
    \mathbf{H}^{(l+1)} = \sigma \left( \hat {\mathbf{A}} \mathbf{H}^{(l)} \mathbf{W}^{(l)} \right),
\end{align}
\begin{align}
    f \left( \mathbf{A}, \mathbf{X} \right) = \sigma \left( \hat{\mathbf{A}} \mathbf{X} \mathbf{W} \right).
\end{align}

To capture the temporal dependence of the sequence data, RNN is widely used. LSTM and GRU use a gate mechanism to resolve the vanishing gradient problem of RNN. Compared to LSTM, GRU has a lighter structure, fewer parameters, and a shorter training time. 
$h_{t-1}$ is the hidden state at time ${t-1}$, the multivariate time series data $\mathbf{X}_t$ is the input data at time $t$, and $h_t$ is the hidden state at time $t$.
%
The update gate $u_t$ determines how much of the current state should be reflected, based on the previous hidden state $h_{t-1}$ and the current input $\mathbf{X}_t$. 
The reset gate $r_t$ controls the extent to which the information from the previous hidden state $h_{t-1}$ should be forgotten. 
The new hidden state $c_t$ is calculated by combining the partially reset information from the previous state, controlled by the reset gate $r_t$, with the current input $\mathbf{X}_t$. 
The final hidden state $h_t$ is determined by a weighted sum of the previous hidden state $h_{t-1}$ and the newly calculated state $c_t$, where the weights are provided by the update gate $u_t$. Gates and hidden states can be determined as 
\begin{align}
    u_t = \sigma \left( W_u \left[ f(\mathbf{A}, \mathbf{X}_t), h_{t-1} \right] + b_u \right),
\end{align}
\begin{align}
    r_t = \sigma \left( W_r \left[ f(\mathbf{A}, \mathbf{X}_t), h_{t-1} \right] + b_r \right),
\end{align}
\begin{align}
    c_t = \tanh \left( W_c \left[ f(\mathbf{A}, \mathbf{X}_t), (r_t \odot h_{t-1}) \right] + b_c \right),
\end{align}
\begin{align}
    h_t = u_t \odot h_{t-1} + (1 - u_t) \odot c_t,
\end{align}
where $W$ and $b$ are weights and biases of the gate. $\odot$ denotes element wise multiplication.

\subsection{Fault Detection} 

T-GCN model captures the spatial-temporal patterns and predicts the next behavior of the model. When the model trains on normal data, it predicts the behavior of normal test data as normal. As abnormal data with defects exhibit different characteristic patterns from normal data, the model fails to make accurate predictions. If the predicted output is different from the actual behavior, the difference can be an indicator representing the signal of the fault in the system. Fault detection approaches typically use the quantitative indicator called anomaly score to detect whether a fault has occurred or not. If the score exceeds a threshold, it can be determined that a failure has occurred. The threshold is generally the maximum MSE of the actual train data and the predicted train data.

\subsection{Fault Severity} 

The fault severity estimation can identify the size or the level of the fault. 
One conventional approach treats fault severity as a discrete variable. Faults are categorized into indices such as minor, moderate, and critical. The fault levels are used as categorical labels that a classification model learns from input features. The model learns the relationship between the input data and these labels to accurately predict fault levels. 

Another approach employs a fault monitoring index, using metrics such as the Mahalanobis Distance or the Kullback-Leibler Divergence\cite{yang2022incipient}. This numerical metric is used to quantitatively assess the severity of detected faults by comparing to the healthy state of the system. A larger index value indicates a more severe fault.

However, these conventional approaches have notable limitations. Discrete indices of fault severity, while simple and intuitive, may not capture the continuous progression of fault conditions. This discrete categorization may lead to inconsistent index transitions and unstable fault estimations when changes occur gradually. In contrast, continuous indices provide a more granular assessment by representing fault severity as a continuously varying metric, allowing for the detection of subtle changes in system behavior and offering improved sensitivity and robustness in fault evaluation.

Additionally, many previous studies rely on predefined fault metrics, including extent of damage, leakage rate, and weight variation. Such reliance requires domain expertise to establish appropriate thresholds and parameters, which can restrict the approach's adaptability and generalizability across different applications.

\section{Robust Fault Analysis and Severity Estimation} \label{method} 

This study proposes a method for robust fault detection and severity estimation, addressing two key challenges: robustly capturing the relationships among variables and managing the variability in fault severity. To address the first challenge, we employ the T-GCN that leverages both spatial and temporal features to effectively capture variable dependencies in multivariate time-series data. For the second challenge, we propose a novel fault severity estimation method that utilizes the statistical characteristics of the anomaly scores.

\subsection{Robust Fault Severity Estimation} 

This section describes the index of fault level for robust fault severity estimation. Statistical characteristics such as mean and standard deviation are used to reduce the fluctuation of the function that indicates the anomaly score. 
This index exclusively uses anomaly scores that exceed a threshold $\tau$ for analysis. Let $S_t$ denote the anomaly score at time $t$. An indicator function representing identified faults is defined as
\begin{align} \label{It}
I_{t} = \begin{cases}
1, ~~\text{if}~ S_t > \tau, \\
0, ~~\text{otherwise}.
\end{cases}
\end{align}
This function assigns a value of 1 when the anomaly score exceeds the threshold, marking the time points at which faults have been detected.
The cumulative count of detected faults up to time $t$ is given by
\begin{align} \label{nt}
n_{t} = \sum_{k=1}^t{I_{k}}.
\end{align}
Here, $n_t$ represents the total number of time instances where a fault was detected. This value serves as a normalization factor in our subsequent calculations of mean, ensuring that the derived metrics are based solely on the occurrences of detected faults.
The statistical properties of the anomaly scores for detected faults can be quantified using the indicator function defined in \eqref{It} and the cumulative fault count provided in \eqref{nt}. Specifically, mean and variance are defined as 
\begin{align} \label{mut}
\mu_{t} = \frac{1}{n_{t}} \sum_{k=1}^{t} {\max (0,S_k - \tau)},
\end{align}
\begin{align} \label{sigmat}
\sigma_t^2 = \frac{1}{n_{t}} \sum_{k=1}^t{(\max(0,S_k -\tau))^2} - \mu_{t}^2.
\end{align}
The index provides a comprehensive measure of both the average deviation and the variability of the anomaly scores during fault conditions.
The index is defined as
\begin{align} \label{Stm}
S_{t,m} = \mu_t + m \sigma_t,
\end{align}
where $m$ is a tunable parameter that governs the relative weight of the variability ($\sigma_t$) in the overall fault index. It can be selected based on the system's specific characteristics.

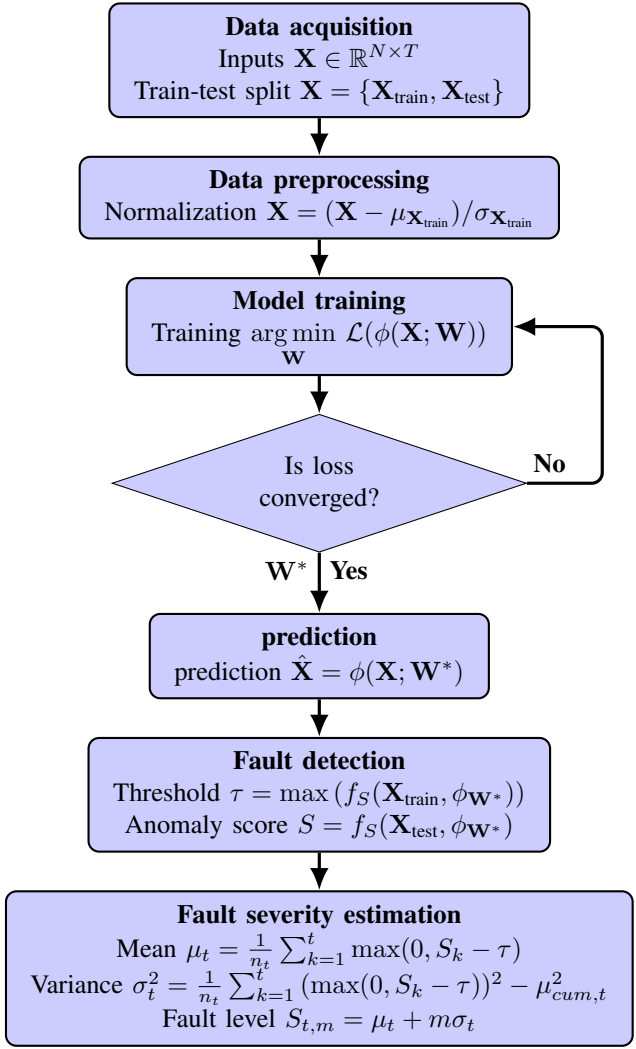
\begin{figure} \begin{minipage}{\columnwidth} \input{tikz/flowchart/flowchart-source.tex} \end{minipage} \caption{Flowchart illustration of the robust fault detection and fault severity estimation algorithm. } \label{alg} \end{figure}



\subsection{Robust Fault Analysis and Severity Estimation Algorithm} 

Fig.\ref{alg} presents a flowchart of our proposed method for robust fault detection and severity estimation in complex systems. By combining deep learning with statistical properties, the algorithm effectively handles multivariate inputs and reduces sudden fluctuations in fault severity. A statistical approach is applied to transform these anomaly scores into fault severity indices, offering a robustly quantified measure of how severe a detected fault may be. The mean of the index represents the tendency of the anomaly score to change. The standard deviation of the index represents the potential range of fluctuation in the anomaly score. 

The detailed explanation is as follows. The algorithm begins by collecting multivariate time-series data from sensors embedded within the system. This data, which consists of several variables measured over time, including rotating speed and vibration magnitude, is divided into training and test sets. The training set predominantly reflects normal operating conditions. To ensure consistent scaling among the variables and facilitate effective model learning, the data is normalized using the mean and standard deviation derived from the training set. Specifically, each variable is transformed as 
\begin{align}
\mathbf{X} = (\mathbf{X} - \mu_{\mathbf{X}_{\text{train}}}) / \sigma_{\mathbf{X}_{\text{train}}},
\end{align}
where $\mu_{\mathbf{X}_{\text{train}}}$ and $\sigma_{\mathbf{X}_{\text{train}}}$ are the mean and standard deviation of the train data. This normalization reduces scale differences among variables, allowing the model to capture underlying patterns more effectively.

 After normalization, the preprocessed data is input into the model to learn normal pattern of the system. The training objective is to find the optimal parameter $\mathbf{W}^*$ that minimizes the loss function as
\begin{align}
\underset{\mathbf{W}}{\arg\min} \ \mathcal{L}(\phi(\mathbf{X}; \mathbf{W})),
\end{align}
where $\mathcal{L}$ is loss function and $\phi$ represents the model’s mapping. Training process continues until the training loss converges. Once training is complete, the optimized parameter $\mathbf{W}^*$ is determined. 
The predicted output $\hat{\mathbf{X}}$ is generated using the optimized parameter and the input data as 
\begin{align}
\hat{\mathbf{X}} = \phi(\mathbf{X}; \mathbf{W}^*).
\end{align}

Once training is complete, the model detects faults by generating anomaly scores. A function that indicates the anomaly score is defined as the mean squared error (MSE) between the predicted output and the actual behavior:
\begin{align}
S = f_S (\mathbf{X}_{\text{test}}, \phi_{\mathbf{W}^*} ),
\end{align}
where $f_S$ is the function for anomaly score and $\phi_{\mathbf{W}^*}$ represents the trained model. To determine whether a test data point corresponds to a fault, it is necessary to establish a threshold for anomaly detection. In this study, the threshold is set as the maximum prediction error observed in the training data as
\begin{align}
\tau = \max \left( f_S (\mathbf{X}_{\text{train}}, \phi_{\mathbf{W}^*} ) \right),
\end{align}
where $\tau$ is the threshold. This approach ensures that the model accounts for a reasonable level of prediction error inherent in normal system behavior. If the anomaly score of a test data point exceeds the threshold, the data is detected as a fault. 

The fault severity index provides a robust measure of fault progression by analyzing only the anomaly scores that exceed a predefined threshold. The mean represents the average amount of detected faults, while the standard deviation quantifies their fluctuation. The severity index is calculated as the sum of the mean and a weighted multiple of the standard deviation, where the parameter adjusts the sensitivity to fault variability. This formulation ensures a stable and adaptive fault severity estimation, making it suitable for complex systems with varying fault behaviors.

A larger value of $m$ increases the weight assigned to variability in the fault index calculation. Consequently, even slight fluctuations in the anomaly scores result in a significantly higher computed fault index. This adjustment ensures that deviations are robustly reflected in the final assessment, yielding a more conservative evaluation. This approach is particularly valuable in safety-critical domains such as nuclear power plants, where even minor anomalies in sensor data may indicate potential operational risks.

\section{Experiments and Results} \label{exp}

To investigate the performance of the proposed method, we conduct fault detection and fault severity estimation for two examples of experiments. The first example uses the IMS bearing dataset, and the second example uses a fanjet electric propulsion experiment. From the result of the evaluation, we verify the applicability of the proposed method.

\subsection{IMS Bearing Dataset} 

\subsubsection{Experimetnal setup} 

IMS bearing dataset is open public data with information on bearing healthy and failure conditions. The dataset consists of three run-to-failure vibration experiments. For the study, the second one of the three tests is adapted. Four bearings are installed on a shaft which is coupled to AC motor. RPM of the motor is kept constant at $2000$ RPM. Four accelerometers are installed on the housing of the bearings. The test consists of $20,480$ points with $1$-second vibration snapshots. The recording interval is $10$ minutes, and the test lasts for $9850$ minutes. Outer race failure occurs in bearing $1$, but there is no information indicating when the failure started. 

\subsubsection{Preprocessing and Model Training} 

This section describes the architecture and training procedure of our deep learning model used for robust fault detection and severity estimation. An explanation of the data preprocessing process for training, the loss function, and the optimization method ia provided. 

The adjacency matrix A is designed to reflect the relationships between variables in each dataset. The matrix is essential for capturing spatial dependencies in multivariate time-series data, allowing the T-GCN model to effectively learn interactions between different signals.
The bearing dataset consists of four bearings, each monitored for vibration signals. Since the physical arrangement of the bearings affects how faults propagate, we construct the adjacency matrix based on their spatial connectivity. Specifically, we assume that only adjacent bearings influence each other, while non-adjacent bearings have no effect at all. Thus, the adjacency matrix is defined as
\begin{align} \label{A_bearing}
\mathbf{A} = \begin{bmatrix}
0 & 1 & 0 & 0 \\
1 & 0 & 1 & 0 \\
0 & 1 & 0 & 1 \\
0 & 0 & 1 & 0
\end{bmatrix}.
\end{align}

The bearing dataset is divided into training, validation, and test sets to best suit their characteristics. The dataset allocates $60\%$ of the data for testing, while the remaining $40\%$ is used for training. From the training data, $20\%$ is set aside for validation. 
Standardization is applied to control the data scale. To analyze time-series data using GRU, the sliding window technique is employed to construct training datasets, with a window size set to $4$. And the size of the batch is set to $32$. 
The T-GCN model described above is utilized. The dimension of the hidden layer is set to $128$ and the number of layers is set to $2$.
The loss function is mean squared error. To minimize the loss function, Adam is used as an optimizer, with a learning rate set to $0.001$. 
The model is trained for $50$ epochs.

\subsubsection{Fault Detection and Fault Severity Estimation}

After training the T-GCN model on normal data, the trained model is used to detect faults and estimate fault severity on test data. The process consists of two main parts: fault detection through anomaly scoring and fault severity estimation based on statistical properties.

\begin{figure}[!t] \vspace*{0.05in} \centering \includegraphics[width=2.5in]{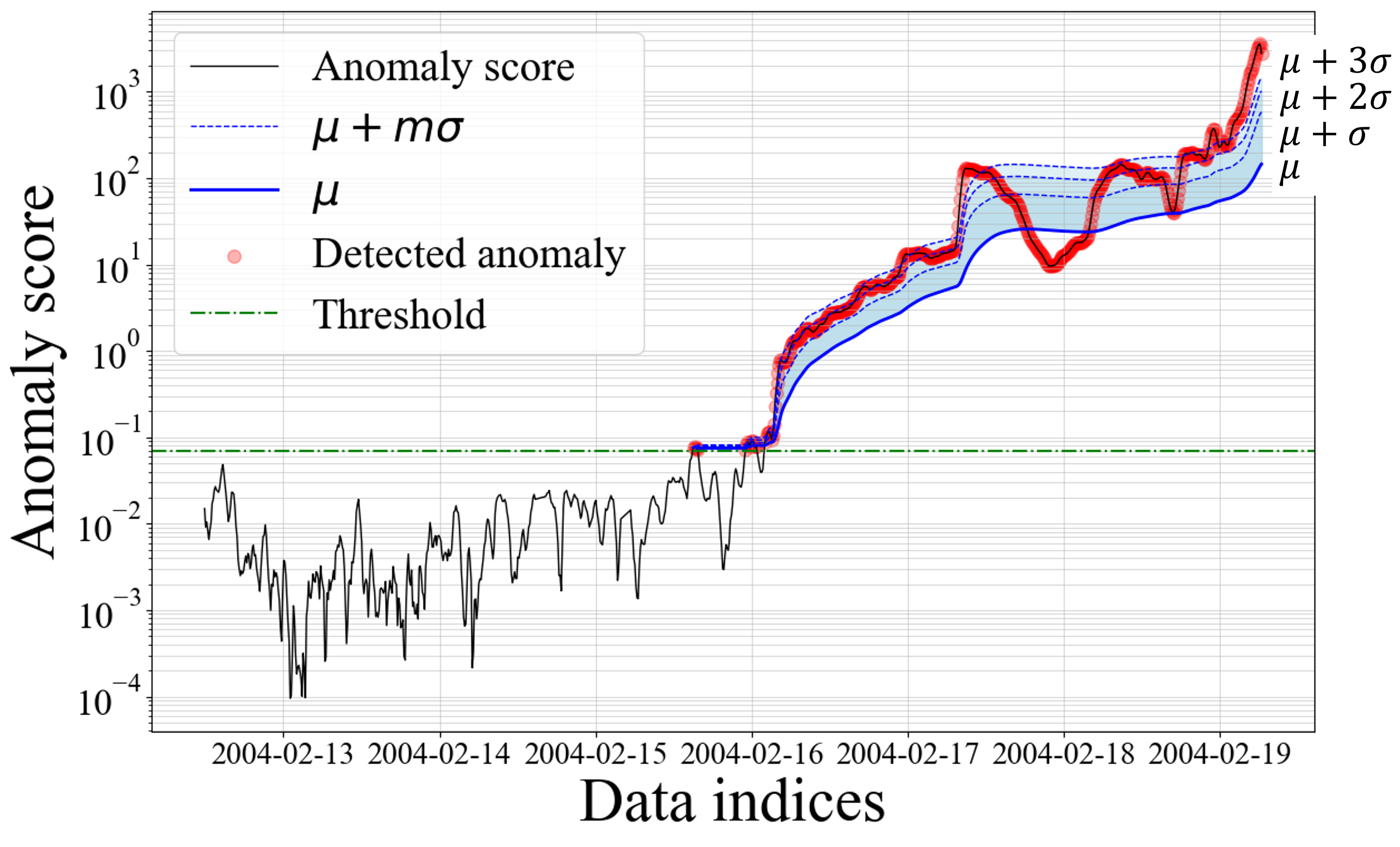} \caption{Fault severity estimation for bearing 1 in the bearing dataset} \label{bearing_1} \end{figure}

\begin{figure}[!t] \vspace*{0.05in} \centering \includegraphics[width=2.5in]{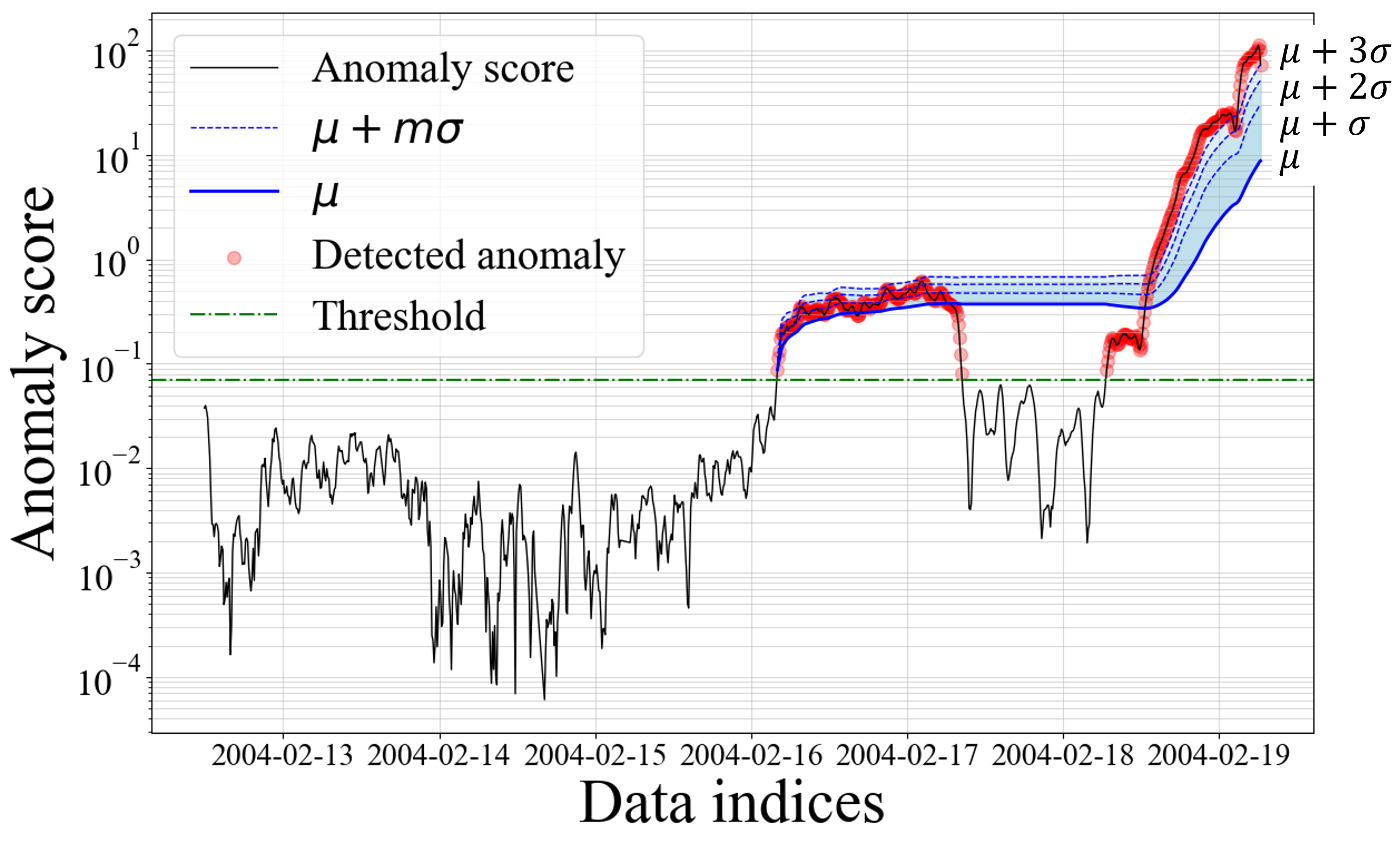} \caption{Fault severity estimation for bearing 2 in the bearing dataset} \label{bearing_2} \end{figure}

\begin{figure}[!t] \vspace*{0.05in} \centering \includegraphics[width=2.5in]{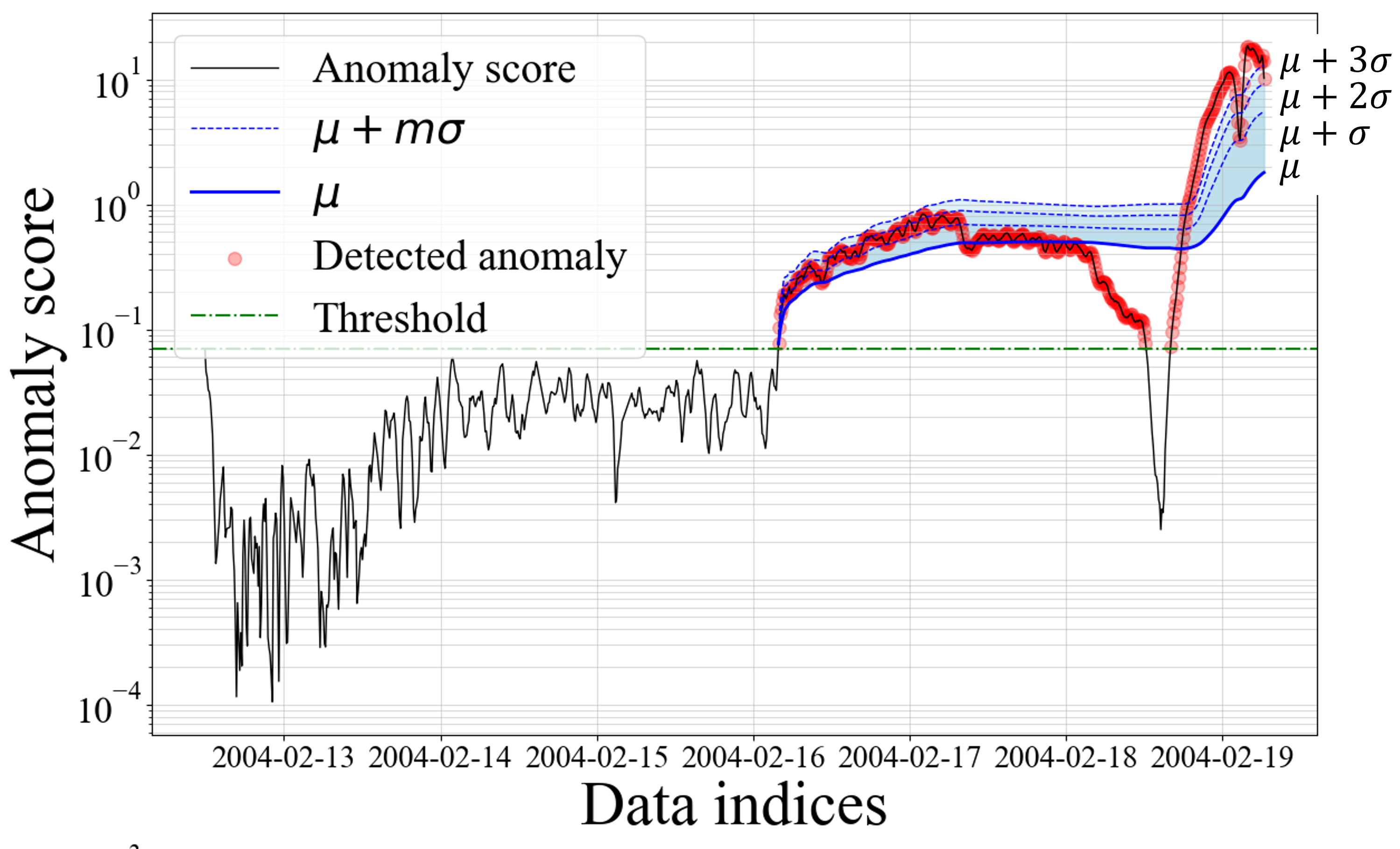} \caption{Fault severity estimation for bearing 3 in the bearing dataset} \label{bearing_3} \end{figure}

\begin{figure}[!t] \vspace*{0.05in} \centering \includegraphics[width=2.5in]{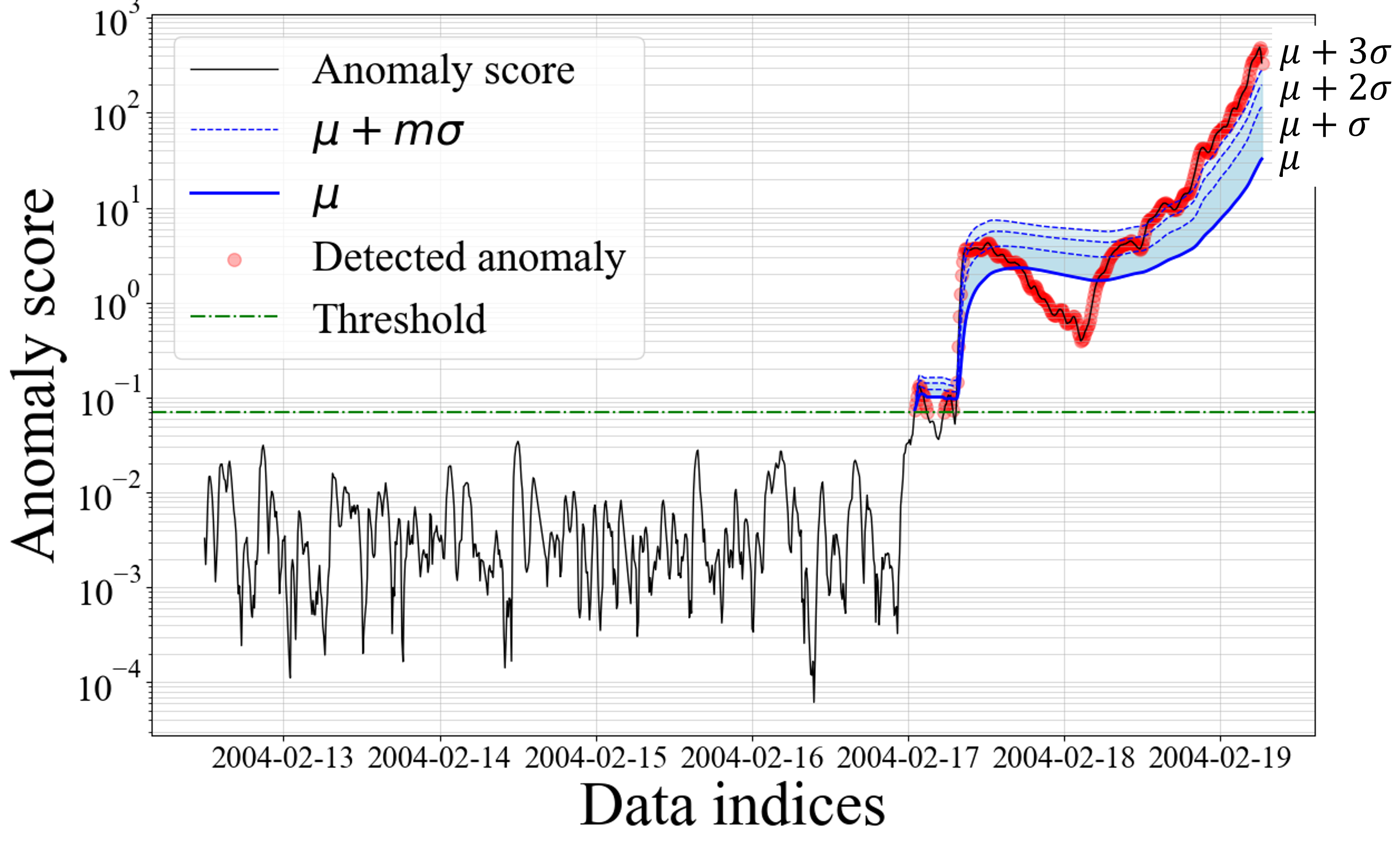} \caption{Fault severity estimation for bearing 4 in the bearing dataset} \label{bearing_4} \end{figure}

Fig. \ref{bearing_1}-\ref{bearing_4} show the anomaly scores, thresholds, and estimated fault severity for bearings $1-4$ in the bearing dataset. The solid black line indicates the anomaly score, while the green horizontal line represents the threshold derived from the training set. Whenever the anomaly score exceeds this threshold, the model detects the data points (red markers) as anomalies. The exceeded score indicates the severity of fault. The solid and dashed blue lines correspond to $\mu$ and $\mu+m\sigma$, respectively, providing a quantitative measure of how the fault severity evolves over time. 

The figures reveal that the anomaly scores exhibit considerable fluctuations, making it challenging to precisely determine the current stage of fault progression. Additionally, the faults are intermittently detected and then subsequently dismissed, leading to inconsistent fault detection. Such repetitive variations in fault detection and severity outcomes can cause confusion when interpreting the true state of the system. 
Therefore, our proposed method is employed to mitigate these issues. By integrating a robust model with statistical properties that stabilize the derived fault severity indices, our approach yields a more reliable and consistent evaluation of system faults.


\subsection{Fanjet Experiment Dataset} 

\subsubsection{Experimetnal setup} 

\begin{figure}[!t] \vspace*{0.05in} \centering \includegraphics[width=2.5in]{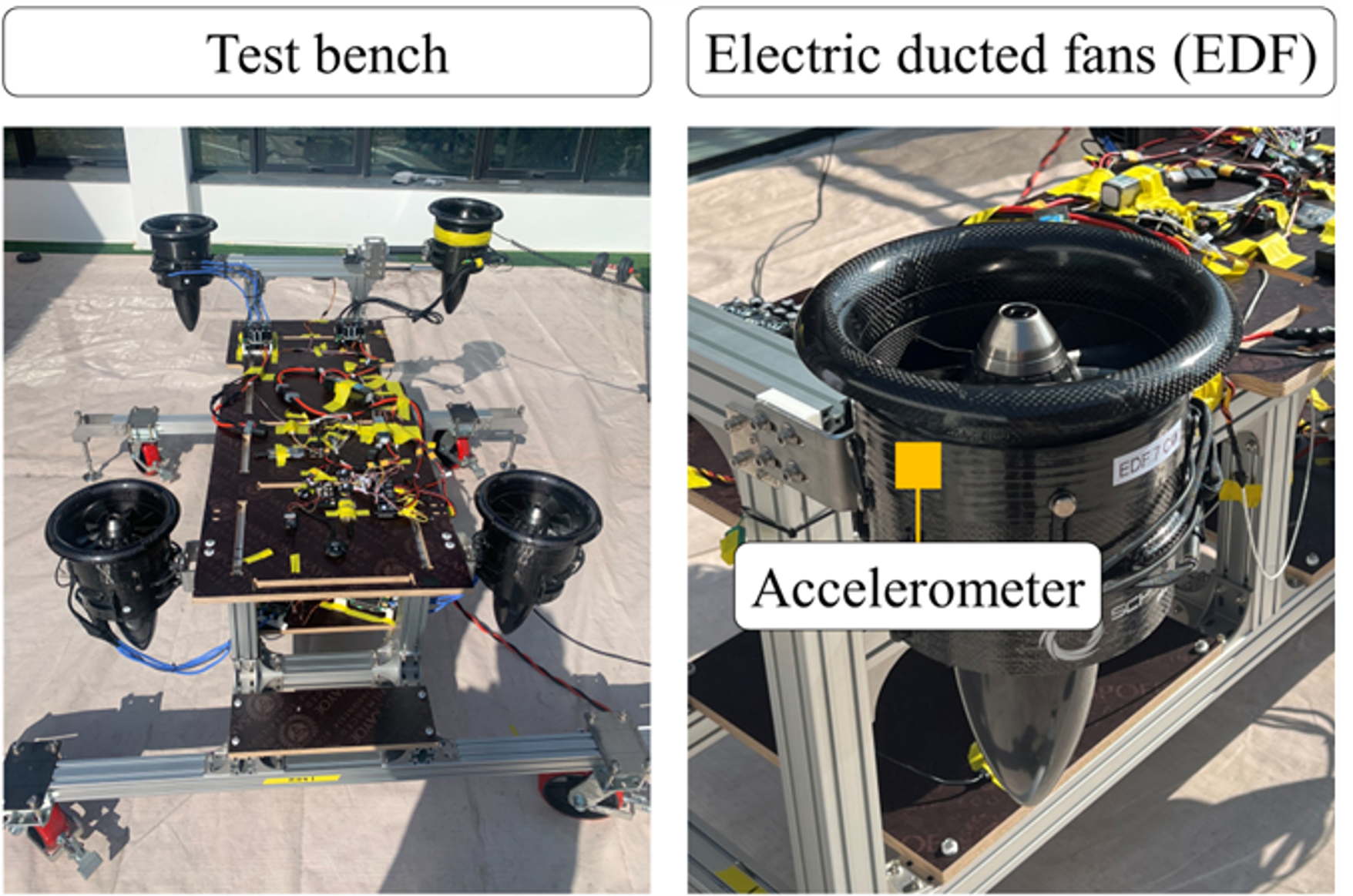} \caption{Test bench of eVTOL propulsion system with four electric ducted fans} \label{testbench} \end{figure}

This experiment is conducted to measure the vibration and rotational speed data of the fanjet. For the experiment, we build a test bench to replicate the propulsion system of an $3$ meter wide eVTOL. The test bench consists of four electric ducted fans (EDF), electric propulsion controller, power supply, Pixhawk4-based control computer, and ground control unit to drive the propulsion system.
The fanjet is attached to a test bench, as shown in Fig.\ref{testbench}. RPM of the EDF is kept constant at around $50$ RPM. An accelerometer is mounted on the case of the test EDF to measure its vibration. The test data consists of root mean square of $1$-second vibration snapshots, and the recording interval is $15$ seconds.

Due to experimental constraints, we are unable to induce an actual fault in the EDF. Instead, we simulate a fault condition to collect fault data. We assume a fault situation where the EDF is loosely attached on the eVTOL, due to issues like loose screws. To replicate this scenario, the EDF is intentionally oscillated from zero to three degrees during vibration measurement, simulating the conditions of a loosely attached system. In other words, the EDF is subjected to a $3$-degree angular oscillation to mimic the vibrations caused by loose mounting in real-world operations. This experimental setup is designed to mimic the vibrations that could occur in real-world operations when the EDF is not securely fixed.

\subsubsection{Preprocessing and Model Training} 

In the fanjet dataset, the primary variables include vibration magnitude and RPM. Unlike the bearing dataset, the fanjet dataset consists of only two variables that are inherently related—RPM fluctuations directly affect vibration characteristics. We define an adjacency matrix as
\begin{align} \label{A_fanjet}
\mathbf{A} = \begin{bmatrix}
0 & 1 \\
1 & 0 
\end{bmatrix}.
\end{align}
The fanjet dataset allocate $40\%$ of the data for testing, while the remaining $60\%$ is used for training. From the training data, $30\%$ is designated for validation.
A window size is set to $4$ and the batch size is set to $4$. 
$2$-layer T-GCN model is utilized and the dimension of the hidden layer is set to $256$. 
The loss function is mean squared error. To minimize the loss function, Adam is used as an optimizer, with a learning rate set to $0.001$. 
The model is trained for $50$ epochs.

\subsubsection{Fault Detection and Fault Severity Estimation}

\begin{figure}[!t] \vspace*{0.05in} \centering \includegraphics[width=2.5in]{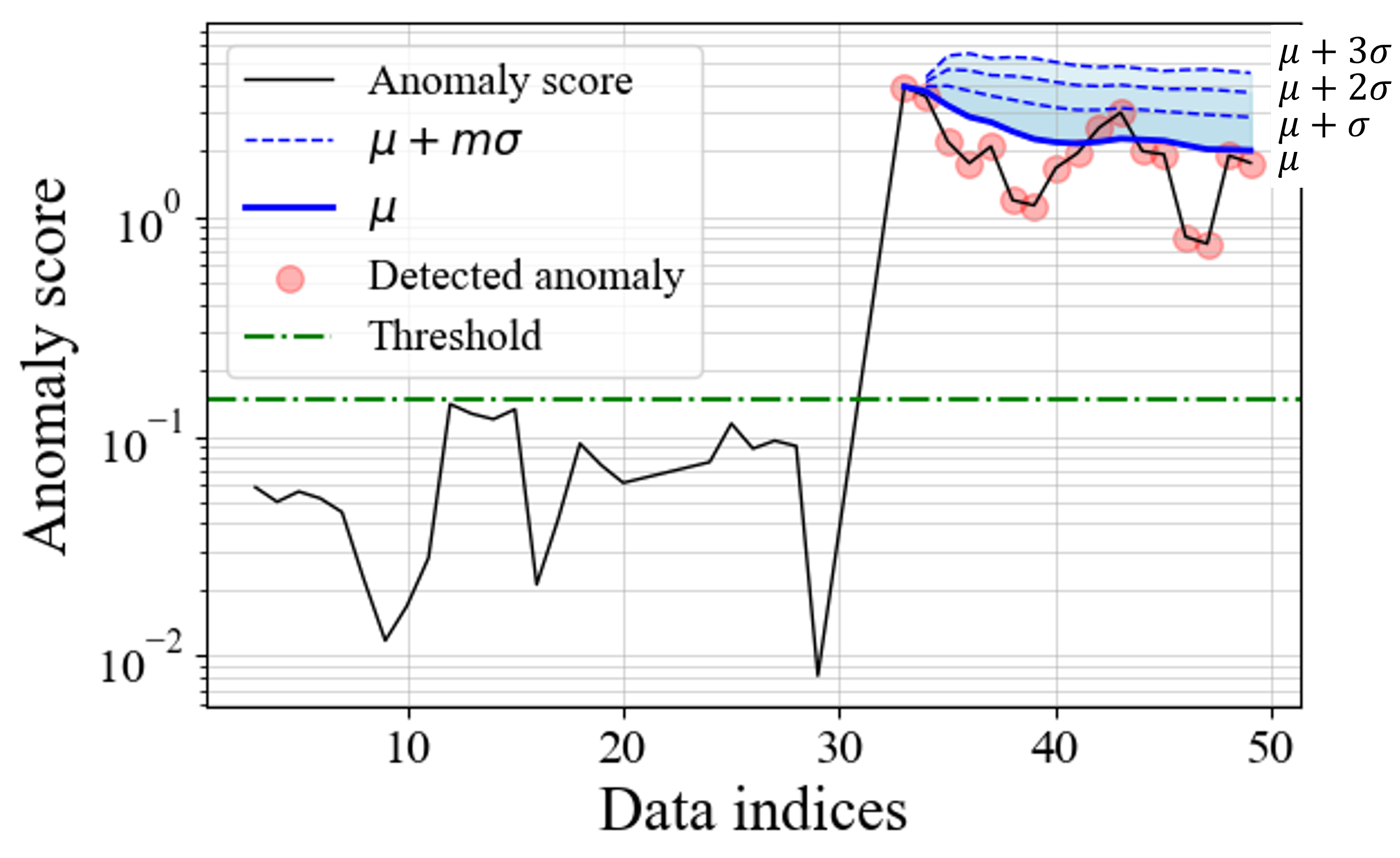} \caption{Fault severity estimation for vibration magnitude in the EDF dataset} \label{fanjet_vib} \end{figure}

\begin{figure}[!t] \vspace*{0.05in} \centering \includegraphics[width=2.5in]{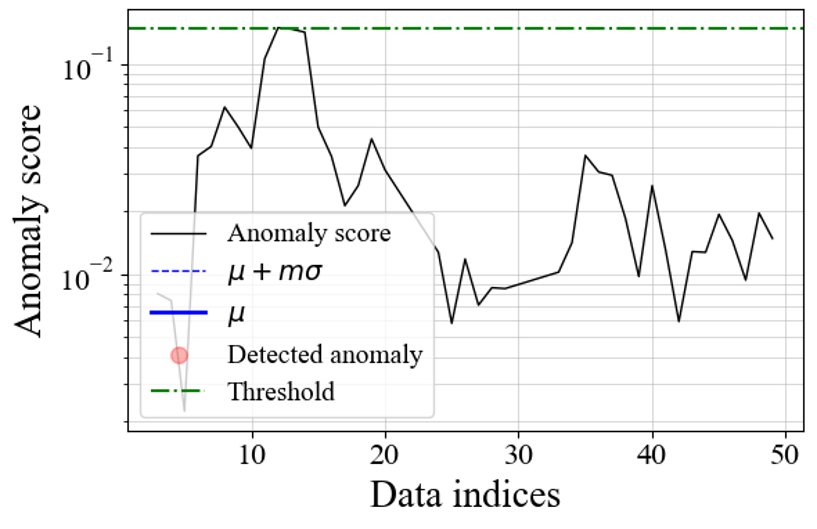} \caption{Fault severity estimation for RPM in the EDF dataset} \label{fanjet_rpm} \end{figure}

Fig. \ref{fanjet_vib} and \ref{fanjet_rpm} illustrate anomaly score results for the fanjet dataset, focusing on vibration magnitude and RPM, respectively. Here, the model also detects anomalies once the anomaly score surpasses the threshold. The fault severity indices, $\mu$ and $\mu+m\sigma$, offering a robust representation of how significantly the system deviate from normal operation.


\section{Conclusion} \label{con}

We developed a method for robust fault detection and fault severity estimation to address two key challenges: capturing relationships in multivariate time-series data and stabilizing fluctuations in fault severity. Our approach integrates the T-GCN model with statistical properties of the anomaly score.
By leveraging the T-GCN model, the proposed method captures the continuity and relationships among variables. 
Additionally, by leveraging the statistical properties of the anomaly score, such as the mean and standard deviation, it effectively reduces sudden fluctuations in fault severity, ensuring a more stable and reliable evaluation.
The index that quantifies the anomaly score addresses two challenges: managing fluctuations in fault severity and reducing the reliance on domain expertise for predefined fault levels. 
By modifying the weight of the standard deviation, the fault index can be adjusted to align with the characteristics of the system.
We evaluated our approach using two datasets: an open bearing dataset and fanjet data obtained from our own experiments. The results indicate that our fault severity index significantly reduces sudden fluctuations and inconsistent fault detection, compared to conventional methods.

The index produced by our approach serves as a critical indicator in system monitoring. The proposed method assists operators in planning maintenance actions, paving the way for improved operational reliability and proactive risk management.
These strengths are particularly beneficial in situations where reliable fault severity is required as well as in cases where establishing predefined fault level is challenging. 
One limitation of our study is that the fanjet dataset was collected under a single operating RPM condition, which restricts the evaluation of our method under varying operational speeds. In future research, experimental data will be collected under various RPM conditions to further enhance the robustness of our approach.


\section*{Acknowledgment} 

This was supported by Korea Institute for Advancement of Technology(KIAT) grant funded by the Korea Government(MOTIE)
(P0017120, The Competency Development Program for Industry Specialist).



\bibliographystyle{IEEEtran}
\bibliography{FDD}
%



\end{document}

%% file: tikz/flowchart/flowchart-source.tex
\usetikzlibrary{shapes.geometric}

\begin{tikzpicture}

    \node (data) [base] {\begin{tabular}{c} \textbf{Data acquisition} \\ Inputs $\mathbf{X} \in \mathbb R ^{N\times T}$ \\ Train-test split $\mathbf{X}=\{\mathbf{X}_{\text{train}}, \mathbf{X}_{\text{test}}\}$ \end{tabular}};

    \node (pre) [base, below=0.5cm of data.south] {\begin{tabular}{c} \textbf{Data preprocessing} \\ Normalization $\mathbf{X} = (\mathbf{X} - \mu_{\mathbf{X}_{\text{train}}}) / \sigma_{\mathbf{X}_{\text{train}}}$ \end{tabular}};

    \node (model) [base, below=0.5cm of pre.south] {\begin{tabular}{c} \textbf{Model training} \\ Training $\underset{\mathbf{W}}{\arg\min} \ \mathcal{L}(\phi(\mathbf{X}; \mathbf{W}))$  \end{tabular}};

    \node (converge) [draw, diamond, aspect=3, below=0.5cm of model.south, fill=blue!20] {\begin{tabular}{c} \text{Is loss} \\ \text{converged?} \end{tabular}};
    
    \node (predict) [base, below=0.8cm of converge.south] {\begin{tabular}{c} \textbf{prediction} \\  prediction $\hat{\mathbf{X}} = \phi(\mathbf{X}; \mathbf{W}^* )$ \end{tabular}};
    
    \node (anomaly) [base, below=0.5cm of predict.south] {\begin{tabular}{c} \textbf{Fault detection} \\ Threshold $\tau = \max \left( f_S (\mathbf{X}_{\text{train}}, \phi_{\mathbf{W}^*} ) \right)$ \\ Anomaly score $S = f_S (\mathbf{X}_{\text{test}}, \phi_{\mathbf{W}^*} )$ \end{tabular}};

    \node (severity) [base, below=0.5cm of anomaly.south] {\begin{tabular}{c} \textbf{Fault severity estimation} \\ Mean $\mu_{t} = \frac{1}{n_{t}} \sum_{k=1}^{t} {\max (0,S_k - \tau)}$ \\ Variance $\sigma_t^2 = \frac{1}{n_{t}} \sum_{k=1}^t{(\max(0,S_k -\tau))^2} - \mu_{cum,t}^2$ \\ Fault level $S_{t,m}=\mu_t+m\sigma_t$ \end{tabular}};

    \draw [connector] (data) -- (pre);
    \draw [connector] (pre) -- (model);
    \draw [connector] (model) -- (converge);
        \draw [connector] (converge.east) -- ++(1,0) node[pos=0.3, above] {\textbf{No}} |- (model);
    \draw [connector] (converge) -- node[pos=0.3, left] {$\mathbf{W}^*$} node[pos=0.3, right] {\textbf{Yes}} (predict);
    \draw [connector] (predict) --(anomaly);
    \draw [connector] (anomaly) -- (severity);
    
\end{tikzpicture}